\documentclass[twocolumn,prl,showpacs,superscriptaddress]{revtex4}
\usepackage{epsf}
\usepackage{color}
\usepackage{verbatim}
\usepackage{hyperref}
\usepackage{graphics}
\usepackage{graphicx}
\usepackage{natbib}
\usepackage{dcolumn}
\usepackage{bm}

\begin{document}
\title{Quantum oscillations of ortho-II high temperature cuprates}
\author{Daniel Podolsky}
\affiliation{Department of Physics, University of Toronto, Toronto,
Ontario M5S 1A7 Canada}
\author{Hae-Young Kee}
\email{hykee@physics.utoronto.ca}
\affiliation{Department of Physics, University of Toronto, Toronto,
Ontario M5S 1A7 Canada and 
School of Physics, Korea Institute for Advanced Study, Seoul 130-722, Korea }
\date{\today}

\begin{abstract}
Motivated by quantum oscillations observed in highly ordered ortho-II 
YBa$_2$Cu$_3$O$_{6.5}$, we study the Fermi surface topology of d-density wave (ddw) or antiferromagnetic (AF) ordering
in the presence of the ortho-II potential.
We find that the electron pocket is unaffected by the presence of the ortho-II potential.  This further strengthens the proposal that
quantum oscillations observed in ortho-II-free YBa$_2$Cu$_4$O$_8$ arise from an electron pocket. On the other hand, the hole pocket topology is sensitive to the ortho-II potential.
We show there exist generically three distinct quantum oscillations
associated with  one electron-like and two hole-like Fermi pockets. 
We compare our results to the quantum oscillations observed in experiments.  We discuss
possible ways to distinguish between ddw and antiferromagnetic orders in a potential single-layer
ortho-II material.
\end{abstract}
\maketitle

{\it Introduction} ---
Quantum oscillations in the magnetization (de Haas-van Alphen) and conductivity (Shubnikov-de Haas)
are powerful tools to probe the Fermi surfaces of complex materials.
However, the search for quantum oscillations in
high temperature cuprates has not been successful until very recently.
The exhibition of clear oscillations were first reported in ortho-II YBa$_2$Cu$_3$O$_{6.51}$\cite{leyraud,jaudet}
and  YBa$_2$Cu$_4$O$_8$\cite{bangura}.
The first observation of Shubnikov-de Haas (SdH) oscillations in ortho-II YBa$_2$Cu$_3$O$_{6.51}$ (YBCO)
with $T_c = 57.5K$ and nominal doping $p_{nom} =0.1$
proves the existence of a closed Fermi surface in the {\it normal} state of the underdoped cuprates.
While the applied magnetic fields are lower than $H_{c2}$, the quantum oscillations and their frequency 
in the mixed state are properties of the normal state.\cite{wasserman} 

The frequency $F$ of $1/B$ oscillations is measured in field units, and is proportional to the 
area $A_k$ enclosed by a closed Fermi surface.
The size of the Fermi surface determined by the oscillation frequency in ortho-II YBCO is
too small to match the nominal doping of $0.1$ --
a frequency of 530 T implies a Fermi surface pocket which
is only 1.9\% of the original Brillouin zone.  
Assuming that there are 4 (2) pockets, this leads to $p=0.152$ (0.076) doping which
is far different from the nominal doping of $0.1$.  In the absence of translational symmetry breaking, the disagreement is even worse, as the area of the Fermi pockets in that case should add up to $1+p$. This would indicate that there must be more than
one type of Fermi pocket -- an observation that is consistent with the presence of quantum oscillations in the
Hall coefficient.  In addition, the fact that the Hall coefficient is negative implies that the charge carrier in at least one of the
Fermi pockets is electron-like rather than hole-like \cite{leboeuf}.

Shortly after the discovery of quantum oscillations, three different proposals have been made.\cite{millis,
chakravarty,chen}
The common aspect of the proposals is that a state with broken translational symmetry  is responsible for
the observed oscillations, but they differ in the precise nature of the broken symmetry.
More recently, an additional oscillatory component, with frequency $\approx 1650$ T 
has been observed in the same sample of ortho-II YBCO.\cite{sebastian} 
This poses a challenge to the proposed order scenarios.\cite{millis,chakravarty,chen,dimov,jia}.   For example, the ddw and AF states produce a hole pocket in addition to the electron Fermi pocket.
However, the frequency associated with the hole pocket is fixed at 970 T by the Luttinger sum rule.\cite{chakravarty} 
It has also been suggested that the quantum oscillations are due to an incommensurate helical order.\cite{jia,sebastian}

In this paper, we offer a phenomenological theory which captures quantum oscillations of the two different
observed frequencies within the ddw order proposal.  
The  key idea of the present work is to take $(\pi,\pi)$ ordering (which can be either ddw or AF order)  and  
the ortho-II potential into account on equal footing.
This leads to new Fermi surface shapes for the hole pocket
while the electron pocket topology is insensitive to the presence of the ortho-II potential 
We generally show that there are three closed Fermi pockets.  They lead to three oscillatory components, associated with one electron
and two hole pockets, that are constrained by the Luttinger sum rule to satisfy $F_\beta+F_\gamma-F_\alpha\approx 1400$ T.  We will also show that quantum oscillations on  {\it a single-layer ortho-II compound} would give  a way to distinguish between ddw and AF orderings.  In addition, we will discuss the angle-resolved
photoemission spectroscopy (ARPES) experiments.

\begin{figure}[htb]
\epsfxsize=10cm
\includegraphics*[angle=0, width=1.0\linewidth, clip]{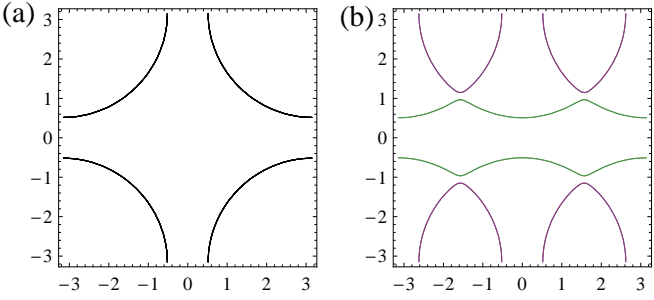}
\caption{The Fermi surface of tight-binding model on a square lattice without (a) and with (b) the ortho-II potential.
We set $t=0.3$, $t^{\prime} = -0.09$, and $t^{\prime \prime} =0.012$, and
$\mu=-0.266$, and the external potential from the ortho-II structure $\lambda=0.025$.
}
\label{fig:orthoII}
\end{figure}

{\it Ortho-II band structure} --- 
The electronic band dispersion on a square lattice is given by 
$\epsilon_{\bf k} = - 2 t (\cos{k_x}+\cos{k_y}) - 4 t^{\prime} \cos{k_x}
\cos{k_y} - 2 t^{\prime \prime} (\cos{2 k_x} + \cos{2 k_y}) - \mu$, 
where we set the lattice spacing $a \equiv 1$, and $t$, $t^{\prime}$, and $t^{\prime \prime}$ are
the nearest, next-nearest, and third-nearest  hopping integrals, respectively.
The highly ordered chains in ortho-II YBCO induce a period-two potential $\lambda$, with ordering vector $(\pi,0)$.  This modifies the band dispersion, which becomes
\begin{eqnarray}
\epsilon^{\pm}_{\bf k} &=& -2 t \cos{k_y} - 2 t^{\prime \prime} (\cos{2 k_x}+\cos{2 k_y})
-\mu \nonumber\\
& & \pm \left( 4 \cos^2{k_x} (t+2 t^{\prime} \cos{k_y})^2 + \lambda^2 \right)^{\frac{1}{2}}, 
\end{eqnarray}
see Fig.~\ref{fig:orthoII}.
We take $t=0.3$, $t^{\prime}=-0.09$, $t^{\prime \prime}
=0.012$ and the external potential $\lambda = 0.025$ \cite{t_footnote}. The hopping integrals 
and the ratio between $\lambda$ and $t$ are similar to those used in Ref. \cite{chakravarty} 
and  Ref. \cite{bascones}, respectively.
The ortho-II phase of YBa$_2$Cu$_3$O$_{6.5}$ is characterized by  alternating empty and filled Cu chains
along $b$-axis which doubles the unit cell in the $a$-direction.\cite{band-structure} 
We will include the bilayer coupling $t_{\perp}$ later to see a pure effect of bilayer coupling.

\begin{figure}[htb]
\epsfxsize=8.5cm
\includegraphics*[angle=0, width=0.95\linewidth, clip]{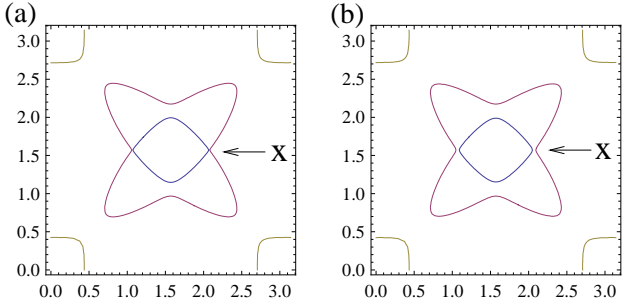}
\caption{The Fermi surface in one quadrant of the original Brillouin zone 
on a single layer for (a) the ddw state ($\Delta_{ddw}=0.02$ and $\mu=-0.266$) and (b) the AF state ($\Delta_{AF}=0.07$ and $\mu=-0.27$).   In both cases, we take $t=0.3$, $t'=-0.09$, $t''=0.012$, and $\lambda=0.025$.
Note that point X is degenerate only for the ddw state.
}
\label{fig:ddwSingle}
\end{figure}

{\it  Fermi surface topology for $(\pi,\pi)$ ordering on a single layer} ---  
The ddw ordering is characterized by an alternating current on a square plaquette\cite{chakravarty2,affleck,palee}, with order parameter
\begin{equation}
\Delta_{ddw} = i \sum_{\bf k} 2\left( \cos{k_x}-\cos{k_y} \right) \langle c^{\dagger}_{\bf k}
c_{{\bf k}+{\bf Q}} \rangle,
\end{equation}
where $Q =(\pi,\pi)$.
Thus the quasiparticle spectrum in the ddw state with the ortho-II structure
can be found by diagonalizing the following $4\times 4$ matrix,
\begin{eqnarray}
H = \sum_{\bf k} \left( \begin{array}{cccc} \epsilon_{\bf k} & \lambda & 0 & \Delta_{{\bf k}} \\
 \lambda  & \epsilon_{{\bf k}+(\pi,0)} & \Delta_{{\bf k}+(\pi,0)} & 0\\
0 & \Delta^*_{{\bf k}+(\pi,0)} & \epsilon_{{\bf k}+{\bf Q}+(\pi,0)} &  \lambda  \\
 \Delta^*_{{\bf k}} & 0 &  \lambda  & \epsilon_{{\bf k}+{\bf Q}} \end{array} \right).
\end{eqnarray}
$H$ is written in the basis $(c_{\bf k}, c_{{\bf k}+(\pi,0)}, 
c_{{\bf k}+{\bf Q}+(\pi,0)}, c_{{\bf k}+{\bf Q}})$, and $\Delta_{\bf k} = i 2 \Delta_{ddw} (\cos{k_x}
-\cos{k_y} ) $. 
Fig. \ref{fig:ddwSingle} shows the Fermi surface at $\mu=-0.266$ which leads to the doping of 10 \% in
the quadrant of the original Brillioun zone. 
We set the ddw order amplitude $\Delta_{ddw} = 0.02$.

Two degenerate bands intersect along the $k_y=\pi/2$ line in Fig.~\ref{fig:ddwSingle}.   However,
as we will show below, the bilayer coupling lifts this degeneracy and opens up a gap between the
two bands.  On the other hand, AF order lifts this degeneracy even on a single layer. We will discuss
this degeneracy in more detail later on, and argue that the degeneracy gives a possible way to distinguish between
ddw and AF order in single-layer compounds.

\begin{figure}[htb]
\epsfxsize=14cm
\includegraphics*[angle=0, width=1.0\linewidth, clip]{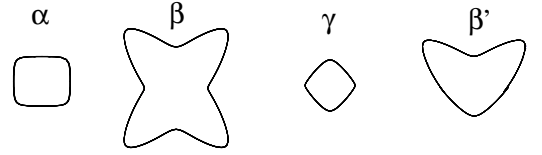}
\vskip -1cm
\caption{The closed Fermi surfaces discussed in this paper.  The associated
frequencies are (for a single layer) $F_\alpha\approx 540$ T, $F_\beta\approx 1560$ T, $F_\gamma\approx 430$ T, and $F_{\beta'}\approx 1000$ T. Note that
$F_{\beta'}$ is the same frequency as the hole pocket in the ddw state of ortho-II free YBCO.\cite{chakravarty}}\label{fig:bands}
\end{figure}

\begin{figure}[htb]
\epsfxsize=4 cm
\includegraphics*[angle=0, width=0.6\linewidth, clip]{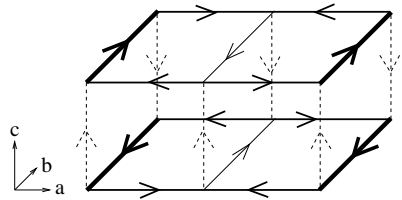}
\caption{Current pattern of a bilayer (current intensity is proportional to arrow thickness).   The CuO chains lie along the $b$ axis, and the current pattern is staggered as we move along $b$. The ddw order is out of phase between the two layers.  The ortho-II potential induces
inter-layer currents, as shown. }
\label{fig:bilayerCurrents}
\end{figure}

{\it Bilayer coupling} --  
The bilayer coupling in YBCO has a dramatic effect.  First of all, hopping
between the layers leads to hybridization of the electronic bands on the two layers, and to splitting of these bands.  In addition, 
in the ddw state,  the bilayer coupling induces a pattern of inter-layer currents.  This induced current
is intimately connected to the ortho-II potential.  It was reported that the ortho-II
potential induces a charge modulation in the plane \cite{bascones}.  Similarly, it can generate a modulation in the current magnitude, such that
the current in the $b$-axis alternates in magnitude along the $a$-axis.  When the bilayer coupling is present, the alternating
current magnitude is accommodated by allowing currents to flow in between the layers, as shown in Fig.~\ref{fig:bilayerCurrents}.

To capture the interlayer currents,  we introduce an additional term to the Hamiltonian,
$2i \tilde{u}( \cos{k_y} c^{\dagger +}_{\bf k} c^{+} _{{\bf k}+(0,\pi)}-\cos{k_y} c^{\dagger -}_{\bf k} c^{-} _{{\bf k}+(0,\pi)}  - c^{\dagger +}_{\bf k} c^{-}_{{\bf k}+(0,\pi)}+ c^{\dagger -}_ {\bf k}c^{+}_{{\bf k}+(0,\pi)} )$, where $\pm$ denotes
the layer index. These terms represent alternating currents in the b-c plane and between two layers,
respectively.   Introducing ${\bf k}_1={\bf k}+(\pi,0)$, ${\bf k}_2={\bf k}+(0,\pi)$, and ${\bf k}_3={\bf k}+{\bf Q}$,
the mean field  Hamiltonian is,
\begin{eqnarray}
 \left( \begin{array}{cccccccc} \epsilon_{\bf k} &  \lambda  &  - u_{\bf k} & \Delta_{{\bf k}} &
t_{\bf k} & 0 & u_0 & 0 \\
 \lambda  & \epsilon_{{\bf k}_1} & \Delta_{{\bf k}_1} & -u_{{\bf k}_1} & 0 & t_{{\bf k}_1}
& 0 &  u_0 \\
-u_{\bf k}^*  & \Delta^*_{{\bf k}_1} & \epsilon_{{\bf k}_2} &  \lambda  & u_0 & 0 & t_{{\bf k}_2} & 0 \\
 \Delta^*_{{\bf k}} & -u_{{\bf k}_1}^* &  \lambda  & \epsilon_{{\bf k}_3} & 0 & u_0 & 0 &
t_{{\bf k}_3} \\ 
t_{{\bf k}} & 0 & u_0^* & 0 & \epsilon_{\bf k} &  \lambda  & u_{\bf k} & \Delta^*_{\bf k} \\
0 & t_{{\bf k}_1} & 0 & u_0^* &  \lambda  & \epsilon_{{\bf k}_1} & \Delta^*_{{\bf k}_1}
& u_{{\bf k}_1} \\
u_0^* & 0 &  t_{{\bf k}_2} & 0 & u_{\bf k}^* & \Delta_{{\bf k}_1} & 
\epsilon_{{\bf k}_2} &  \lambda  \\
0 & u_0^* & 0 & t_{{\bf k}_3} & \Delta_{\bf k} & u_{{\bf k}_1}^* &  \lambda  & \epsilon_{{\bf k}_3} 
\end{array} \right).
\end{eqnarray}
Here $t_{\bf k} = t_\perp ( \cos{k_x} - \cos{k_y})^2/4$ is the bilayer coupling,\cite{bilayer}
$u_{\bf k} =-2i\tilde{u}\cos{k_y} $, and  $\tilde{u}={\mathcal O}(\lambda \Delta_{ddw}/t)$.

Here we consider the case where the order parameter between layers is out-of-phase.  While in-phase ordering is also possible, it is not favored by the antiferromagnetic coupling
between the layers.  The two cases yield different neutron scattering signals, with peaks at
momentum $(\pi,\pi,\pi)$ for the out-of-phase case, and $(\pi,\pi,0)$ in-phase.

\begin{figure}[htb]
\epsfxsize=4cm
\includegraphics*[angle=0, width=0.62\linewidth, clip]{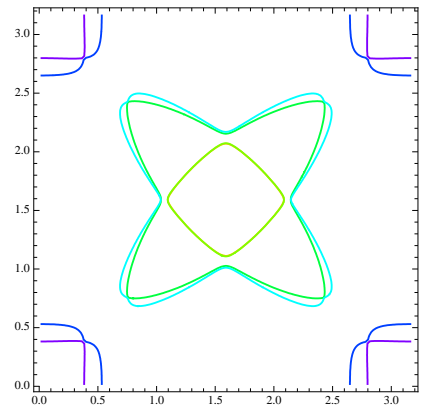}
\caption{Fermi surface in one quadrant of the original Brillouin zone 
in the ddw state with bilayer coupling, with $\tilde{u}=-0.008$ and $t_{\perp} = 0.025$.
Other parameters are $t=0.3$, $t'=-0.09$, $t''=0.012$, $\lambda=0.025$, $\Delta_{ddw}=0.02$, and $\mu=-0.266$.}
\label{fig:bilayerFS}
\end{figure}

The Fermi surface in the bilayer system is shown in Fig. \ref{fig:bilayerFS}.  
The main effect of the bilayer coupling is to change the hole Fermi surface topology -- the degeneracy along the $k_y=\pi/2$ direction has been lifted by the inter-layer currents.
This yields three types of bands ($\alpha$, $\beta$, and $\gamma$ in Fig.~\ref{fig:bands}). 
In addition, each one of these bands is split into two bands due to hybridization.  However, since the ddw is out of phase on the two layers, the quasiparticles 
are not eigenstates under exchange of the two layers. As a consequence, the resulting bands are not well-separated, 
and they nearly intersect.  This is true independently of the presence or absence of the ortho-II potential.  For instance, in the absence of the ortho-II potential,  the quasiparticle energy is given by
\begin{equation}
\frac{\epsilon_{\bf k} +\epsilon_{{\bf k}+{\bf Q}}}{2} \pm 
\frac{1}{2} \sqrt{ (\epsilon_{\bf k} - \epsilon_{{\bf k}+{\bf Q}} \pm 2 t_{\bf k} )^2 + 4 \Delta_{\bf k}^2}
\nonumber
\end{equation}
Hence, the bilayer coupling does not lead to significant splitting of the frequencies.
On the other hand, when the ddw is in phase on the two layers,
the quaisparticles are eigenstates under exchange of the two layers, so
the bilayer bands are well-separated due to the bilayer coupling:
\begin{equation}
\frac{\epsilon_{\bf k} +\epsilon_{{\bf k}+{\bf Q}}}{2} \pm 
\frac{1}{2} \sqrt{ (\epsilon_{\bf k} - \epsilon_{{\bf k}+{\bf Q}} )^2 + 4 \Delta_{\bf k}^2} \pm t_{\bf k}
\nonumber
\end{equation}
A weak dispersion along the $c$ axis has a different effect, which 
splits $F_\alpha$ into two slightly different frequencies.

{\it Quantum oscillations} ---  
Thus far we have shown that $(\pi,\pi)$ ordering on ortho-II YBa$_2$Cu$_3$O$_{6.51}$ gives rise to three closed 
Fermi pockets (Fig.~\ref{fig:bands}).  The electron pocket $\alpha$ is centered about $(0,0)$ and is rectangular.  The hole
pocket $\beta$ ($\gamma$) appears near $(\pi/2,\pi/2)$ and is flower-shaped (diamond-shaped).  Each pocket gives rise
to its own characteristic frequency of quantum oscillations.
The frequency $F$ of $1/B$ oscillations is measured in field units, and is proportional to the area $A_k$ enclosed by the pockets, 
 $F= \phi_0 A_k/(4\pi^2)$, where $\phi_0=hc/e$ is the flux quantum.  In principle, each band can give rise to higher harmonics, 
 but typically the amplitude of these oscillations is very small.

The frequencies $F_{\alpha,\beta,\gamma}$ are constrained by the Luttinger sum rule.   Consider a single layer.  Then, the density of carriers
as a fraction of Cu sites is $p=A_k ab/(2\pi)^2$, where $a$ and $b$ are the lattice constants.  Since there is one
pocket of each type in the reduced Brillouin zone, the total hole doping is $p=p_\beta+p_\gamma-p_\alpha$.  Note that for each pocket $i$, $F_i=(\phi_0/2ab) p_i$.
Thus, the doping $p~10\%$ corresponds to the constraint $F_\beta+F_\gamma-F_\alpha\approx 1400$ T.   Experimental uncertainty in the hole doping $p$ can lead to 
slightly different values for the constraint, {\it e.g.} $F_\beta+F_\gamma-F_\alpha\approx 1540$ T for $p=11\%$.

For the Fermi surface in Fig.~\ref{fig:bilayerFS}, the resulting frequencies
are $F_\alpha\approx 530$ T, $F_\beta\approx 1580$ T, and $F_\gamma\approx 410$ T.   These numbers correspond to
a specific choice of parameters, as shown in the captions of  Figs.~\ref{fig:orthoII}, \ref{fig:ddwSingle}, 
and \ref{fig:bilayerFS}.    The effect of different parameters is as follows: (1) The ortho-II potential $\lambda$ controls the ratio between $F_\beta$ and $F_\gamma$, while leaving $F_\alpha$ unaffected.
(2) The parameters $\Delta_{ddw}$, $t''$, and $\mu$ determine the ratio between $F_\alpha$ and $F_\beta$.   (3)
 The bilayer coupling $t_\perp$ induces a small gap between the $\beta$ and $\gamma$ bands, but its precise value does not affect
the oscillation frequencies appreciably.  
In fact, due to constraints placed by the Luttinger sum rule and by the experimental observation
of strong quantum oscillations with frequency $F_\alpha\approx 530$ T, parameters cannot
be markedly different from our choice.   We find that, in order to satisfy these constraints, the combination of $F_\beta$ and $F_\gamma$, $F_{\beta}+F_{\gamma}$ should be  $1900T$.
Hence, these frequencies are robust features in our model, unless the nominal doping is different from 10\%.\cite{liang}

Experimentally, the dominant quantum oscillations are at a frequency $\approx 530$ T \cite{leyraud,jaudet,bangura}, which we attribute to 
the $\alpha$ pocket.  Recently S. Sebastian {\it et al.}  found evidence of an additional, large
pocket in dHvA oscillation measurements on a single crystal of underdoped YBa$_2$Cu$_3$O$_{6.5}$  \cite{sebastian}.
The new oscillatory component is 30 times smaller in magnitude
than the $\alpha$ contribution, and perhaps for this reason was not seen in earlier dHvA measurements \cite{jaudet}.
The reported frequency, 1650 T, is consistent with our $\beta$ band.  A strong prediction from our model is the presence of a third
oscillation arising from the $\gamma$ band, with frequency $F_\gamma \sim 1900-1650 =250 $T.  
%
However,  a small $\gamma$ band oscillation frequency indicates that a larger window of magnetic field
is required to detect a few oscillations to confirm its existence.
Furthermore, the $\gamma$ band occupies a similar region of the Brillouin zone as the $\beta$ band, it is likely that quasiparticles in these two bands have similarly large scattering rates.\cite{chakravarty}
If these effects  dominate, then the 
the oscillation from $\gamma$ band would be difficult to observe.

We note that a large Fermi surface pocket is hard to obtain from a stripe order, since the stripe ordering
tends to generate smaller pockets and open Fermi surfaces rather than large pockets.\cite{millis}  
It was suggested that a large pocket can be found when one takes a single wavevector
of  the either incommensurate spin density wave or 
incommensurate orbital current order at a wavevector of  $(\pi(1-2 \delta), \pi)$ \cite{sebastian}.
The possibility of incommensurate orbital currents is discussed in \cite{chakravarty,dimov,chakravarty2,kee}, and we will
address the relevance of various incommensurate orderings to the observation of
quantum oscillations in the near future.\cite{ic-oc}.  Note that, in the case of YBa$_2$Cu$_4$O$_8$, where the ortho-II potential is absent, the electron
pocket is only slightly modified, $F_\alpha\approx 660$ T -- a change that can be attributed to the different doping $p\sim 12 \%$ -- while the hole pocket should have a frequency $F_{\beta'}\approx 1170$ T \cite{chakravarty}.

{\it Degeneracy in single layer ortho-II potential}---
Now let us proceed to show how the degneracy in 
single layer ortho-II materials can be used to distinguish the two different $(\pi,\pi)$ orderings.
For the ddw order, two bands have a Fermi surface crossing at point X in Fig.~\ref{fig:ddwSingle}(a).  
This degeneracy holds provided that the single-layer system {\it (i)} has an electronic dispersion with
mirror symmetry about a plane perpendicular to the CuO chains, 
\begin{equation}
\epsilon_{(k_x,k_y)} =  \epsilon_{(k_x,-k_y)},
\label{degeneracy-1}
\end{equation}
and  the order parameter satisfies 
\begin{equation}
\Delta_{(k_x+\pi,\pi/2)}=-\Delta_{(k_x,\pi/2)}.
\label{degeneracy-2}
\end{equation}
Note that the last requirement distinguishes the ddw from the AF state -- in the AF state, $\Delta_{AF}$ is momentum independent, and the
degeneracy is lifted.  On the other hand, the degeneracy is maintained for other ddw states, such as the $d_{xy}+id_{x^2-y^2}$ state of Ref.~\cite{yakovenko}.
Thus, quantum oscillations with an ortho-II potential could tell the difference between  
ddw and AF states.  In the AF state, there would be $\alpha$, $\beta$, and $\gamma$ bands, even in the single layer case.
On the other hand, for the ddw in a single layer, the $\beta$ and $\gamma$ bands are replaced by $\beta'$ bands, shown in Fig.~\ref{fig:bands}, with
frequency $F_{\beta'} \approx 1000$ T.  Thus, it would be desirable to design a single-layer material with ortho-II potential, as this would allow to distinguish
between AF and ddw states.  The degeneracy is protected even in the presence of electronic interlayer hopping in layered materials.  
Note that if the ddw currents are staggered between adjacent layers, the ordering
wave-vector is $(\pi,\pi,\pi)$, 
and there are two extremal orbits at $k_z=0$ and $\pi/2$. 
The degeneracy exists at $k_z=\pi/2$ extremal orbit, and
the $\beta'$ oscillations will be still seen. On the other hand, if the ddw current pattern does not 
alternate along the $z$-axis, the above Eqs.~(\ref{degeneracy-1}) and (\ref{degeneracy-2}) are trivially satisfied,
and the interlayer hopping has no effect on the degeneracy.

{\it Discussion and Summary} --- 
Another prediction of our theory is that, at large magnetic fields, the $\beta'$ frequency should also be seen in both the AF and the ddw state in bilayer ortho-II YBCO.  At large fields,
magnetic breakdown of the small gap opened at point X leads to two $\beta'$ bands, instead of a $\beta$ and a $\gamma$ band.\cite{blount,falicov}
For both AF and ddw cases, the gap that opens at point X is of order $(\Delta/t)\lambda$, and therefore the field at which the magnetic breakdown
occurs will be similar for both cases.  The magnetic breakdown will be discussed in detail elsewhere.\cite{ic-oc}

Angle resolved photoemission spectroscopy measurements of 
K-deposited ortho-II YBCO \cite{damascelli}  and of Na$_{2-x}$Ca$_x$Cu$_2$O$_2$Cl$_2$ with 10\% doping \cite{shen} display so-called Fermi arcs, with
the intensity of the spectral weight  concentrated near the node positions
$(\pm \pi/2,\pm \pi/2)$.   
This is at odds with the observation of
closed Fermi pockets inferred from the quantum oscillations measurements.  However, in cases where the Fermi pockets are formed by states with
translational symmetry breaking, it may be difficult for ARPES to see the full shape of the pockets\cite{borisenko}, since the intensity of ARPES is higher along the original (unfolded) quasiparticle dispersion.
However, the relationship between ARPES and quantum oscillations remains to be understood.

In summary, we investigate the Fermi surface topology of ortho-II YBCO.  We find that 
ddw and AF orders lead to a Fermi surface reconstruction
in which three distinct closed Fermi pockets are generated.   Our analysis shows that in a system with $(\pi,\pi)$ ordering, a qualitative difference in the observed quantum oscillations in
ortho-II and non-ortho-II YBCO arises naturally, while the presence of an electron pocket is common.  This calls for a non-trivial check of the $(\pi,\pi)$ ordering scenario in YBCO,
where three frequencies, $F_\alpha$, $F_\beta$, and $F_\gamma$, should be seen in ortho-II YBCO, whereas only two frequencies, $F_\alpha$ and $F_{\beta'}$, would be seen in non-ortho-II materials.
We also propose a way to distiguish AF and ddw orders in a single
layer ortho-II material in the context of quantum oscillations. 


{\it Acknowledgements} 
We are grateful to S. Kivelson for useful discussions.
This work is supported by NSERC of Canada, Canadian Institute for Advanced Research, and Canada Research Chair.

\end{document}